\documentclass[aps,prl,floatfix,nofootinbib,superscriptaddress,twocolumn]{revtex4-1}
\usepackage{graphicx,subfigure}
\usepackage{amsmath,amssymb,amsfonts}
\usepackage{bm}
\usepackage{color}
\usepackage{comment}
\usepackage{slashed}

\def \be {\begin{equation} }
\def \ee {\end{equation}}

\def \bem {\begin{multline}}
\def \eem {\end{multline}}

\def \bes {\begin{subequations} }
\def \ees {\end{subequations}}

\newcommand{\Eq}[1]{Eq.(\ref{#1})}

\begin{document}

\title{The chiral anomaly, Berry's phase and chiral kinetic theory, from world-lines in quantum field theory}
\author{Niklas Mueller}
\email{n.mueller@thphys.uni-heidelberg.de}
\affiliation{Institut f\"{u}r Theoretische Physik, Universit\"{a}t Heidelberg, Philosophenweg 16, 69120 Heidelberg, Germany}
\author{Raju Venugopalan}
\email{raju@bnl.gov}
\affiliation{Physics Department, Brookhaven National Laboratory, Bldg. 510A, Upton, NY 11973, USA}

\date{ \today}

\begin{abstract}
We outline a novel chiral kinetic theory framework for systematic computations of the Chiral Magnetic Effect (CME) in ultrarelativistic heavy-ion collisions. The real part of the fermion determinant in the
QCD effective action is expressed as a supersymmetric world-line action of spinning, colored, Grassmanian point particles in background gauge fields, with equations of motion that are covariant generalizations of
the Bargmann-Michel-Telegdi and Wong equations. Berry's phase is obtained in a consistent non-relativistic adiabatic limit. The chiral anomaly, in contrast, arises from the phase of the fermion determinant;
its topological properties are therefore  distinct from those of the Berry phase. We show that the imaginary contribution to the fermion determinant too can be expressed as a point particle world-line path integral and derive the corresponding anomalous axial vector current. Our results can be used to derive a covariant relativistic chiral kinetic theory including the effects of topological fluctuations that has overlap with classical-statistical simulations of the CME at early times and anomalous hydrodynamics at late times. 
\end{abstract}
\maketitle

The possibility that topological sphaleron transitions can be identified in heavy-ion collision experiments has aroused great interest.
Besides the information they provide about the non-perturbative real time dynamics of the QCD vacuum, sphaleron transitions are conjectured to play a role in electroweak baryogenesis~\cite{Riotto:1999yt,Cohen:1993nk}.
A striking manifestation of the role of topology is the Chiral Magnetic Effect (CME) where, as a consequence of the chiral anomaly, an induced current is generated in the direction of the external magnetic
field~\cite{Kharzeev:2007jp,Fukushima:2008xe}.  Whether such an effect is seen in heavy-ion collisions is still unclear and is a focus of experimental research in the field~\cite{Kharzeev:2015znc,Skokov:2016yrj}.
We note that the CME has been observed in condensed matter systems~\cite{Li:2014bha}.

For the CME to be large enough to be observed, it must be generated at the earliest times in the heavy-ion collision where the magnetic fields are very large initially before dying off rapidly~\cite{Skokov:2009qp,Deng:2012pc}.
First principles weak coupling computations of sphaleron transitions~\cite{Mace:2016svc} in the non-equilibrium Glasma matter produced indicate that the sphaleron transition rate is signficantly
larger~\cite{footnote1} %
%\footnote{The computations are for over-occupied SU(2) gauge fields in a fixed box. More realistic geometries and $N_c=3$ will not modify this qualitative conclusion.}
than the corresponding equilibrium
rate~\cite{Moore:2010jd}. Because the occupancies of gluons in the Glasma are large, classical-statistical simulations can be employed~\cite{Tanji:2016dka} to compute {\it ab initio}, in a sphaleron background, and
in the presence of external magnetic fields~\cite{Mueller:2016ven,Mace:2016shq}, the development of a chiral magnetic, and accompanying ``chiral separation", wave~\cite{Son:2004tq,Kharzeev:2010gd}.

Detailed thermalization scenarios however suggest that thermalization occurs at parametrically later times~\cite{Berges:2013eia,Kurkela:2015qoa}. Because the gluon occupancy is of order unity or lower in this regime, kinetic theory 
provides the appropriate description of quark-gluon dynamics and the classical-statistical simulations of the CME must therefore be matched to this framework. Likewise, at the later times when the quark-gluon matter is strongly coupled, the kinetic description of chiral currents must be matched to anomalous hydrodynamics~\cite{Hongo:2013cqa,Hirono:2014oda}. 

There has been a considerable amount of recent work on chiral kinetic theory, and of the role therein of the well known Berry phase~\cite{Berry:1984jv} and of the chiral anomaly~\cite{Son:2012wh,Stephanov:2012ki,Son:2012zy,Chen:2012ca,
Chen:2013iga,Chen:2014cla,Basar:2013qia,Stone:2013sga,Dwivedi:2013dea,
Stone:2014fja,Manuel:2014dza,Manuel:2015zpa,Sun:2016nig,Hidaka:2016yjf}. However, despite significant progress, much work remains to complete a first principles derivation of a relativistically covariant kinetic theory from QCD (or even QED). This is especially true for the treatment of topological fluctuations which the chiral current will experience as it traverses the fireball. 
Further, as first observed by Fujikawa, there is often a conflation of the topology of Berry's phase with that of the chiral anomaly~\cite{Deguchi:2005pc,Fujikawa:2005tv,Fujikawa:2005cn}.

In this letter, we will outline the elements of a first principles world-line derivation of relativistic chiral kinetic theory. The real part of the fermion determinant can be expressed exactly in terms of the supersymmetric quantum mechanical path integral for point particle world-lines~\cite{Strassler:1992zr,Mondragon:1995va,Mondragon:1995ab,Hernandez:2008db,JalilianMarian:1999xt,Schubert:2001he,Bastianelli:2006rx,
Corradini:2015tik}, where the internal spin and color degrees of freedom are expressed in terms of Grassmann variables~\cite{Berezin:1976eg,Ohnuki:1978jv}. We will demonstrate how Berry's phase arises in a specific non-relativistic and adiabatic limit of the corresponding Euler-Lagrange equations of motion for the spinning and colored Grassmanian fields~\cite{Balachandran:1976ya,Balachandran:1977ub,Barducci:1976xq,Barducci:1982yw,Brink:1976uf}.
In contrast, explicit identification of a Berry phase is not relevant in ultrarelativistic contexts; the semi-classical phase space description of world-line trajectories provides all the essential elements in the construction of a covariant chiral kinetic theory. 

The fermion determinant has a relative phase that is well known to be related to the physics of the chiral anomaly~\cite{Fujikawa:1979ay,AlvarezGaume:1983ig,AlvarezGaume:1983at,Polyakov-book}. We will here adapt a trick due to D'Hoker and Gagn{\'e}~\cite{D'Hoker:1995ax,
D'Hoker:1995bj} to express this phase as a point particle path integral nearly identical to the one obtained for the real part. The only difference is that the gauge fields are multiplied by a parameter that regulates chiral symmetry breaking. We will outline how the anomaly arises in this context and clearly  demonstrate that its origin is distinct from the Berry phase. The vector and axial vector current are treated on the same footing and are essential elements in constructing a chiral kinetic theory. Many details of our computations, and several new results, will be given in an accompanying longer paper~\cite{long-paper}. 

We begin with the Euclidean action for massless fermion fields in the background of a vector ($A$) and an auxilliary Abelian axial-vector ($B$) field \cite{footnote2} 
\begin{align}
S_F[A,B]=\int d^4x\;\bar{\psi}\left(i\slashed{\partial}+\slashed{A}+\gamma_5\slashed{B} \right)\psi,\label{eq:classicalaction}
\end{align}
and allow the fermion fields to transform under internal (gauge) symmetry. Performing the path integral over the fermion fields, one obtains the effective action
\begin{align}
-W[A,B]=\log\det(\theta)\qquad\theta= i\slashed{\partial}+{\slashed{A}}+\gamma_5{\slashed{B}}\label{eq:effectiveaction}
\end{align}
The determinant of $\theta$ carries a relative phase. One can therefore formally split \Eq{eq:effectiveaction} into real and imaginary parts,
\begin{align}
W[A,B]=W_\mathbb{R}[A,B]+iW_\mathbb{I}[A,B]\label{eq:realimagpartseff} \, .
\end{align}
The real part of the effective action can expressed as 
\begin{align}
W_\mathbb{R}=-\frac{1}{2}\log\det\left(\theta^\dagger\theta\right) \, .
\end{align}
As shown in \cite{D'Hoker:1995ax,D'Hoker:1995bj}, this can be rewritten as 
\begin{align}
W_\mathbb{R}=-\frac{1}{8}\log\det(\tilde{\Sigma}^2)=-\frac{1}{8}\text{Tr}\log(\tilde{\Sigma}^2)\, .
\label{eq:effactREAL} 
\end{align}
Here $\tilde{\Sigma}^2$ is a sixteen dimensional matrix given by 
\begin{equation}
\tilde{\Sigma}^2=(p-\mathcal{A})^2\;\mathbb{I}_8+\frac{i}{2}\Gamma_\mu\Gamma_\nu F_{\mu\nu}[\mathcal{A}] \,,
\label{eq:Sigma-real}
\end{equation}
 with $\mathbb{I}_8$ the 8-dimensional 
identity matrix and 
\begin{align}
\Gamma_\mu=
\begin{pmatrix}
0 & \gamma_\mu \\
\gamma_\mu  & 0 
\end{pmatrix},\quad
\Gamma_5=
\begin{pmatrix}
0 & \gamma_5 \\
\gamma_5  & 0 
\end{pmatrix},\quad
\Gamma_6=
\begin{pmatrix}
0 & i \mathbb{I}_4 \\
-i \mathbb{I}_4  & 0 
\end{pmatrix},
\end{align}
are $8\times 8$ dimensional gamma matrices. We further define a $\Gamma_7$ matrix anti-commuting with all other elements of the algebra, 
\begin{align}
\Gamma_7=-i\prod\limits_{A=1}^6\Gamma_A=
\begin{pmatrix}
\mathbb{I}_4 & 0 \\
0   & -\mathbb{I}_4 
\end{pmatrix}\,.
\end{align}
The gauge fields in \Eq{eq:effactREAL} that appear explicitly and in the field-strength tensor $F_{\mu\nu}[\mathcal{A}]$ can be split into left-right chiral structures with the $2\times 2$ dimensional matrix form \cite{footnote3},
%\footnote{The $8\times 8$ Gamma matrices in $\tilde{\Sigma}^2$ should be understood 
%as multiplying the diagonal terms of the $2\times 2$ matrices formed by the gauge fields.}
\begin{align}
\mathcal{A}=\begin{pmatrix}
{A}+{B} & 0\\
0 & {A}-{B}
\end{pmatrix}.\label{eq:internalfieldrep}
\end{align}
$\tilde{\Sigma}^2$ admits a manifestly positive definite heat-kernel regularization. Therefore using Schwinger's proper-time scheme, the real part of the effective action can be rewritten as 
\begin{align}
W_\mathbb{R}=\frac{1}{8}\int\limits_0^\infty\frac{dT}{T}\;\text{Tr}_{16}\;e^{-\frac{\mathcal{E}}{2}T\tilde{\Sigma}^2}\, ,
\label{eq:tr}
\end{align}
where $\mathcal{E}$ is the einbein, to be discussed further later. 

This 16-dimensional representation of $\tilde{\Sigma}^2$ is useful because it is conveniently cast into a path integral in terms of Grassmanian variables. These variables are eigenvalues of coherent states of creation/annihilation operators that generate finite dimensional representations of the internal symmetries of the theory~\cite{Berezin:1976eg,Ohnuki:1978jv}. 
One obtains after some algebra,
\begin{align}
W_\mathbb{R}=&\frac{1}{8}\int\limits_0^\infty\frac{dT}{T}\mathcal{N}(T)\int\limits_P\mathcal{D}x\int
\limits_{AP}\mathcal{D}\psi \mathcal{D}\lambda^\dagger \mathcal{D}\lambda\; \mathcal{J}(\lambda^\dagger\lambda)\nonumber\\&\times\left( e^{-\int\limits_0^Td\tau\;\mathcal{L}_{L}(\tau)}+e^{-\int\limits_0^Td\tau\;\mathcal{L}_{R}(\tau)}\right)\,.
\label{eq:real-eff_action}
\end{align}
Some details of this derivation are given in \cite{long-paper}. The point particle Lagrangian for left/right chiralities is 
\begin{align}\label{eq:lagrangianexplicit}
&\mathcal{L}_{L/R}(\tau)=\frac{\dot{x}^2}{2\mathcal{E}}+\frac{1}{2}\psi_a\dot{\psi}_a +\lambda^\dagger\dot{\lambda}\nonumber\\&-\lambda^\dagger\Big[i\dot{x}_\mu(A\pm B)_\mu-\frac{i\mathcal{E}}{2}\psi_\mu\psi_\nu F_{\mu\nu}[A\pm B]\Big]\lambda\,,
\end{align}
with the normalization $\mathcal{N}(T)=\int\mathcal{D}p\; e^{-\frac{\mathcal{E}}{2}\int\limits_0^Td\tau\;p^2(\tau)}$. 
Here  $\psi_a=\sqrt{2}\,\langle \psi|\Gamma_a|\psi\rangle$, with $a=1,\cdots,6$ are Grassmann variables, which are defined over a real vector space, where $|\psi\rangle$ represents a coherent state basis of the Clifford algebra. Likewise, $\lambda^\dagger$ and $\lambda$ are independent Grassmanian eigenvalues respectively of creation and annihilation Fermion operators that generate finite dimensional representation of $SU(N_c)$, where $N_c$ is the number of colors. 
The factor $\mathcal{J}(\lambda^\dagger\lambda) = (\frac{\pi}{T})^{N_c} \sum_\phi \exp[ i\phi (\lambda^\dagger \lambda + N_c/2 -1)]$ is required to project intermediate states in the path integral on to coherent states with unit occupancy~\cite{D'Hoker:1995bj}. The labels $P$ and $AP$ denote periodic and anti-periodic boundary conditions for the configuration space and Grassmanian variables respectively. Henceforth, the Abelian reduction of \Eq{eq:real-eff_action} will be sufficient for our purposes; the extension of our discussion to color degrees of freedom is straightforward~\cite{Balachandran:1976ya,Balachandran:1977ub,Barducci:1976xq}.

Varying the real part of the effective action with respect to the vector gauge field gives the vector current,
\begin{align}
\label{eq:vector-current}
&\langle j_\mu^V(y)\rangle=\frac{\delta\Gamma_\mathbb{R}}{\delta A_\mu(y)}=-\frac{i}{8}\int\limits_0^\infty\frac{dT}{T}\mathcal{N}\int\limits_P\mathcal{D}x\int\limits_{AP}\mathcal{D}\psi\, j_{\mu}^{V,cl}
 \nonumber\\&\times \Bigg( e^{-\int\limits_0^Td\tau\;\mathcal{L}
_L(\tau)}+ e^{-\int\limits_0^Td\tau\;\mathcal{L}_R(\tau)}\Bigg) \,,
\end{align}
with $j_{\mu}^{V,cl}(y)= \int_0^Td\tau\;[\mathcal{E}\psi_\nu\psi_\mu\partial_\nu-\dot{x}_\mu]\delta^4\big(x(\tau)-y\big)$. 
This satisfies both $\partial_\mu  j_\mu^{V,cl} = 0$ and $\partial_\mu \langle j_\mu^V\rangle = 0$. 

The imaginary relative phase in the effective action can be written as 
\begin{align}
W_\mathbb{I}=-\frac{1}{2}\arg\det[\Omega],\qquad
\Omega=\begin{pmatrix}
0 &\theta\\
\theta & 0 
\end{pmatrix},
\end{align}
where $\theta$ is given in \Eq{eq:effectiveaction} and the matrix $\Omega$ is 
\begin{align}
\Omega=\Gamma_\mu(p_\mu-A_\mu)-i\Gamma_7\Gamma_\mu\Gamma_5\Gamma_6 B_\mu\,.
\end{align}
The D'Hoker and Gagn{\'e} ~\cite{D'Hoker:1995ax,D'Hoker:1995bj} trick consists of introducing a parameter that regulates chiral symmetry breaking --- distinct from those employed 
previously~\cite{Polyakov-book} --- to write $W_\mathbb{I}$ as 
\begin{align}\label{eq:heatkernelimag}
W_\mathbb{I}=\frac{i\mathcal{E}}{64}\int\limits_{-1}^1 d\alpha\int\limits_0^\infty dT\;\text{Tr}\left\{ \hat{M}e^{-\frac{\mathcal{E}}{2}T\tilde{\Sigma}^2_{(\alpha)}}\right\},
\end{align}
with a trace insertion 
\begin{align}\label{eq:wlinsertiondef}
\hat{M}=\Gamma_7\,\left( 2\Gamma_5\Gamma_6[\partial_\mu,B_\mu]+[\Gamma_\mu,\Gamma_\nu]\{\partial_\mu,B_\nu \}\Gamma_5\Gamma_6 \right) \mathbb{I}_{2}\,,
\end{align}
that is linear in the axial-vector field and diagonal in the two dimensional field representation space introduced in \Eq{eq:internalfieldrep}. $\tilde{\Sigma}^2_{(\alpha)}$ is identical to the expression in \Eq{eq:Sigma-real},
with $B\rightarrow \alpha\,B$, where $\alpha$ breaks chiral symmetry explicitly for $\alpha\neq \pm 1$.  

This form of the phase of the fermion determinant is useful because it has a heat-kernel structure that can be computed, in a manner identical to the real part, using Grassmanian path integrals~\cite{Berezin:1976eg,Ohnuki:1978jv}. 
The path integral representation of $W_\mathbb{I}$ is given in \cite{long-paper}; it is further shown explicitly there that this representation gives the well-known anomaly relation 
\begin{align}\label{eq:finalanomaly}
\partial_\mu \langle j^5_\mu(y)\rangle\equiv
\partial_\mu\frac{i\delta W_\mathbb{I}}{\delta B_\mu(y)}\Big|_{B=0}=-\frac{1}{16\pi^2}\epsilon^{\mu\nu\rho\sigma}F_{\mu\nu}(y)F_{\rho\sigma}(y)\, .
\end{align}

We will now show how Berry's phase arises from taking a non-relativistic and adiabatic limit of the real part of the effective action $W_\mathbb{R}$. 
Our starting point \cite{footnote4} %
is the world-line Lagrangian in \Eq{eq:lagrangianexplicit} continued to 
Minkowskian metric ($g=\text{diag}[-,+,+,+]$). We proceed by introducing Lagrange multipliers in \Eq{eq:lagrangianexplicit} to i) impose the mass-shell constraint and ii) to project out 
unphysical spin  degrees of freedom for both massless and massive point particles. % Requiring the point-particle action be invariant with respect to these, and employing the Euler-Lagrange 
%equations  %
%\footnote{A detailed discussion of the constrained Hamilton dynamics of world-lines can be found in \cite{long-paper}.}
After imposing all constraints, and eliminating thereby unphysical degrees of freedom~\cite{footnote5}, the Lagrangian can be written as $S= \int_0^{T} d\tau \mathcal{L}$ (setting $\tau=ct\sqrt{1-\mathbf{v}^2/c^2}$, 
with $x^\mu = (ct,\vec{x})$, defining $z=\sqrt{-\dot{x}^2}$ and putting the auxilliary field $B=0$),
\begin{align}
\label{eq:NRLagrangean1}
&\mathcal{L}= -\frac{m_R \,c \,z}{2}\left( 1+\frac{m^2}{m_R^2}\right)+\frac{i}{2}\left(\boldsymbol{\psi} \dot{\boldsymbol{\psi}}- \psi_0\dot{\psi}_0 \right) \nonumber \\
&+\frac{\dot{x}_\mu A^\mu(x)}{c} - \frac{iz}{m_R\,c}\,\psi^0 F_{0i}\psi^i- \frac{iz}{2m_R\,c}\,\psi^i F_{ij}\psi^j \, ,
\end{align}
where $m_R^2=m^2+i\psi^\mu F_{\mu\nu} \psi^\nu/c^2$ is the effective mass for the spinning world-line. The spin three-vector can be defined as $S^i = -\frac{i}{2}\epsilon^{ijk} \psi^j\psi^k$, where $\epsilon^{ijk}$ is the 
Levi-Civita symbol. Likewise, the magnetic field $B^i = \frac{1}{2}\epsilon^{ijk} F^{jk}$ and the electric field $E^i=F^{0i}$. The corresponding equations of motion are the covariant form of the Bargmann-Michel-Telegdi equations~\cite{Bargmann:1959gz} for spinning particles (and likewise, Wong equations for colored
particles~\cite{Wong:1970fu})~\cite{Balachandran:1976ya,Balachandran:1977ub,Barducci:1976xq,Barducci:1982yw,Brink:1976uf}. The last two terms in \Eq{eq:NRLagrangean1} are respectively,  
\begin{align}
-i\psi^0 F_{0i}\psi^i=\frac{\boldsymbol{S}\cdot(\boldsymbol{\pi}\times\boldsymbol{E})}{c\pi^0}\,;\,
-\frac{i}{2}\psi^i F_{ij}\psi^j=\boldsymbol{S}\cdot\boldsymbol{B} \, .
\end{align}
Here $\pi^\mu \equiv p^\mu-A^\mu$, where $p^\mu$ is the canonical momentum. To take the non-relativistic limit, we expand the effective mass as 
$m_R \approx m\,(1+X)$ with $X=-(\boldsymbol{S}\cdot(\boldsymbol{\pi}\times\boldsymbol{E})/[c\pi^0]+\boldsymbol{S}\cdot\boldsymbol{B})/(2\,m^2 c^2)$. Observing that $X\propto (v/c)^2$, where $\vec{v}$ is the three-velocity of the spinning world-line, one can expand \Eq{eq:NRLagrangean1} to obtain 
\begin{align}
\label{eq:NRLagrangean2}
\mathcal{L}_{NR}&=-mc^2+\frac{1}{2}m \boldsymbol{v}^2+\frac{i}{2}\left(\boldsymbol{\psi} \dot{\boldsymbol{\psi}}- \psi_0\dot{\psi}_0 \right) - A^0 +\frac{\boldsymbol{v}}{c}\cdot\boldsymbol{A}\nonumber \\
&+\frac{ \boldsymbol{S}\cdot(\left[ {\boldsymbol{v}/c}-{\boldsymbol{A}/(mc^2)}\right]\times\boldsymbol{E})}{mc}+\frac{\boldsymbol{S}\cdot\boldsymbol{B}}{m} + O\Bigg(\frac{v^3}{c^3}\Bigg)\,.
\end{align}
Compactly expressing the non-relativistic action as $S=\int d t\big( \boldsymbol{p}\cdot\dot{\boldsymbol{x}}+\frac{i}{2}\boldsymbol{\psi}\cdot\dot{\boldsymbol{\psi}}-H \big)$, the corresponding Hamiltonian is \cite{footnotepsi0}
\begin{align}\label{eq:NRhamilt}
H\equiv mc^2&+\frac{\left(\boldsymbol{p}-\frac{\boldsymbol{A}}{c}\right)^2}{2m}+A^0(x)\nonumber\\
&-\frac{ \boldsymbol{S}\cdot(\left[ {\boldsymbol{v}/c}-{\boldsymbol{A}/(mc^2)}\right]\times\boldsymbol{E})}{2mc}-\frac{\boldsymbol{B}\cdot\boldsymbol{S}}{m}\,.
\end{align}
Expressed in this form, the non-relativistic point particle Hamiltonian is familiar~\cite{Sakurai}; the penultimate term is the spin-orbit interaction energy from Thomas precession, while the final term is the Larmor interaction energy. In atomic physics their combined effect is of course to reduce the Larmor energy by the famous ``Thomas~1/2". 

In the following, we will show how the system described by \Eq{eq:NRhamilt} contains, in an adiabatic approximation, a Berry phase; in this limit, it has the monopole form postulated in \cite{Son:2012wh,Stephanov:2012ki,Son:2012zy,Chen:2013iga}. To recover the expressions in \cite{Son:2012wh,Stephanov:2012ki} we re-quantize the spin, 
by promoting the spin (phase-space) variables $\psi$ to the Hilbert space operators $\psi_i\rightarrow\sqrt{\frac{\hbar}{2}}\sigma_i\equiv\hat{\psi}_i$ and $S_i\rightarrow \frac{\hbar}{2}\sigma_i\equiv \hat{S}_k$, where $\sigma$ are the Pauli matrices and hats indicate operators. Further, to describe the finite phase space of  Grasmannian variables $\psi$, we define the two dimensional Hilbert space for a spin-$1/2$ particle at every point in phase space $(\boldsymbol{p},\boldsymbol{x})$ by the eigenstates 
$|\psi^\pm\rangle=|\psi^\pm(\boldsymbol{p})\rangle$.  Defining $\boldsymbol{n}=\frac{\boldsymbol{p}}{|\boldsymbol{p}|}\equiv (\sin\theta\cos\phi,\sin\theta\sin\phi,\cos\theta)$, one has two choices
\begin{align}\label{eq:monoplebundle}
|\psi_+^{(1)}(\boldsymbol{p})\rangle&=\frac{N}{2}\left(1+\boldsymbol{n}\cdot\boldsymbol{\sigma} \right)
\begin{pmatrix}
1\\
0
\end{pmatrix}=\begin{pmatrix}
\cos\frac{\theta}{2}\\
e^{i\phi}\sin\frac{\theta}{2}
\end{pmatrix}\\\label{eq:monoplebundle2}
|\psi_+^{(2)}(\boldsymbol{p})\rangle&=\frac{N}{2}\left(1+\boldsymbol{n}\cdot\boldsymbol{\sigma} \right)
\begin{pmatrix}
0\\
1
\end{pmatrix}=\begin{pmatrix}
e^{-i\phi}\cos\frac{\theta}{2}\\
\sin\frac{\theta}{2}
\end{pmatrix},
\end{align}
for the ``spin up" $+$ basis vectors (where $N$ is a normalization factor) and similarly for the ``spin down" basis vectors (see also \cite{Stone:1985av}).
The two choices of basis vectors are not defined globally for all $\boldsymbol{p}$ with \Eq{eq:monoplebundle} (\Eq{eq:monoplebundle2}) ill defined for the south (north) pole  for $\theta=\pi\;(0)$.
One set can however be used for the northern hemisphere and the other for the southern one, and are related as $|\psi_+^{(1)}(\boldsymbol{p})\rangle=e^{i\phi}|\psi_+^{(2)}(\boldsymbol{p})\rangle$ \cite{Stone:1985av}.

These basis states allow us to derive a path integral formulation in the adiabatic limit of the theory defined by \Eq{eq:NRhamilt}. 
The transition amplitude for the Hamiltonian operator corresponding to \Eq{eq:NRhamilt} from an initial state $|\psi^+(\mathbf{p}_i)\rangle$ at time $t_i$ to the state with momentum $\mathbf{p}_f$ at finite time $t_f$ is 
\begin{align}\label{eq:pathintegraladiabatic}
T(\mathbf{p}_f,\mathbf{p}_i,+)\equiv\langle \mathbf{p}_f,\psi^+(\mathbf{p}_f)| e^{-i\hat{H}(t_f-t_i)}|\mathbf{p}_i,\psi^{+}(\mathbf{p}_i)\rangle\,.
\end{align}
The construction of the path integral for this amplitude requires insertions of complete sets of  intermediate states satisfying
$\mathbb{I}=\int d^3x_k\; | \mathbf{x}_k\rangle\langle \mathbf{x}_k| =\int d^3p_k\; | \mathbf{p}_k\rangle\langle \mathbf{p}_k|$, as well as one for the two dimensional spin-Hilbert space: $\mathbb{I}_2=|\psi^+\rangle\langle\psi^+|+|\psi^-\rangle\langle\psi^-|$. The adiabatic approximation corresponds to $\frac{\boldsymbol{B}\cdot\boldsymbol{S}}{2m}\approx 0$.
Therefore in this limit we can neglect the second term $|\psi^-\rangle\langle\psi^-|$, thereby constraining the dynamical spin degrees of freedom.

The transition matrix element can thus be written as 
%\begin{align}
%&T(\mathbf{p}_f,\mathbf{p}_i,+)=\int\prod\limits_{k=1}^{N-1} \prod\limits_{k^\prime=1}^{N} \frac{d^3 p_k d^3x_{k^\prime}}{(2\pi)^3}\nonumber \\
%& \times e^{-i\mathbf{x}_k^\prime\cdot(\mathbf{p}_k-\mathbf{p}_{k-1})-i H\Delta}\langle \psi^+(\mathbf{p}_k)|\psi^+(\mathbf{p}_{k-1})\rangle\,,
%&T(\mathbf{p}_f,\mathbf{p}_i,+)=\int d^3x_N\; e^{-i\mathbf{x}_N\cdot(\mathbf{p}_N-\mathbf{p}_{N-1})-i H_N\Delta}\langle \psi^+(\mathbf{p}_N)|\psi^+(\mathbf{p}_{N-1})\rangle\nonumber\\
%& \times\int\prod\limits_{k=1}^{N-1} \frac{d^3 p_k d^3x_{k}}{(2\pi)^3}\,e^{-i\mathbf{x}_k\cdot(\mathbf{p}_k-\mathbf{p}_{k-1})-i H_k\Delta}\langle \psi^+(\mathbf{p}_k)|\psi^+(\mathbf{p}_{k-1})\rangle\,,
\begin{align}\label{eq:pathintdisc}
&T(\mathbf{p}_f,\mathbf{p}_i,+)=\int\left( \prod\limits_{k=1}^{N-1} d^3p_k \right)\left( \prod\limits_{l=1}^{N} d^3x_l \right)\nonumber\\&\times\prod\limits_{j=1}^N \frac{1}{(2\pi)^3}e^{-i\mathbf{x}_j\cdot(\mathbf{p}_j-\mathbf{p}_{j-1})-i H_j\Delta}\langle \psi^+(\mathbf{p}_j)|\psi^+(\mathbf{p}_{j-1})\rangle\,,
\end{align}
where $\Delta\equiv (t_f-t_i)/N$ and $H_j$ is \Eq{eq:NRhamilt} evaluated at $(\boldsymbol{x}_j,\boldsymbol{p}_j)$. Taylor expanding $|\psi^+(\mathbf{p}_{j-1})\rangle=\left\{1+[\mathbf{p}_j-\mathbf{p}_{j-1}]\cdot\boldsymbol{\nabla_p} \right\}|\psi^+(\mathbf{p}_{j})\rangle \nonumber+ \cdots$,
it is straightforward to show in the continuum limit that one obtains Berry's phase, 
\begin{align}
\prod\limits_{j=1}^N\langle \psi^+(\boldsymbol{p}_j)|\psi^+(\boldsymbol{p}_{j-1})\rangle\rightarrow \exp{\Big( i\int dt \; \dot{\mathbf{p}}\cdot\boldsymbol{\mathcal{A}}(\boldsymbol{p}) \Big)}\, .
\end{align}
 where $\boldsymbol{\mathcal{A}}(\boldsymbol{p})\equiv -i \langle \psi^+(\boldsymbol{p})|\boldsymbol{\nabla}_p| \psi^+(\boldsymbol{p})\rangle$ is the Berry connection. 
The final expression for the path integral is 
\begin{align}\label{eq:Berrypath}
T(\mathbf{p}_f,\mathbf{p}_i,+)=\int \mathcal{D}x\mathcal{D}p\;\exp{\Big(i \int dt\;\Big[\dot{\mathbf{x}}\cdot\mathbf{p}-\tilde {H}\Big] \Big)} \, ,
\end{align}
with ${\tilde H} = mc^2+\frac{(\boldsymbol{p}-\boldsymbol{A}/c)^2}{2m}+A^0(x)-\dot{\mathbf{p}}\cdot\boldsymbol{\mathcal{A}}(\boldsymbol{p})$.

\Eq{eq:Berrypath} is closely related to a similar formulation in \cite{Son:2012wh,Stephanov:2012ki,Son:2012zy,Chen:2013iga}. 

We note a few crucial points in our derivation and interpretation for systems where the chemical potential is much smaller than the rest mass.
Firstly, as we showed, the structure of Berry's phase is restricted to the non-relativistic adiabatic limit where the Larmor interaction energy is much smaller than the rest energy. It is ill-defined in the massless case albeit the spin basis vectors in \Eq{eq:monoplebundle} look similar to Weyl spinors.
An exception holds for massless systems with a large chemical potential; the latter in that case takes over the role of the mass~\cite{long-paper}.  
Our derivation further makes it clear that the topological structure of Berry's phase \cite{Simon:1983mh} is distinct from that of the chiral anomaly.
The former arises from real part of the QED/QCD effective action while the latter can be traced to its relative phase \cite{footnote6}.

Since the CME dynamics in heavy-ion collisions is relativistic, and far from adiabatic, kinetic theory constructions that explicitly incorporate the Berry phase along the lines of \Eq{eq:Berrypath} are insufficient for this case \cite{footnote7}.
Our world-line expressions in \Eq{eq:vector-current}  for the vector current, and for the axial vector current $\langle j^5_\mu(y)\rangle\equiv \langle \frac{i\delta W_\mathbb{I}}{\delta B_\mu(y)}\rangle\Big|_{B=0}$ in \Eq{eq:finalanomaly} are key ingredients in a real-time many-body world-line kinetic theory. While Berry's phase is not relevant, the effects of the axial anomaly are transparently introduced through $ j^5_\mu(y)$.

The real-time formulation~\cite{Mathur:1993tp}  of a pseudo-classical kinetic theory from a world-line action was worked out for spinless colored particles in \cite{JalilianMarian:1999xt}; the non-Abelian Boltzmann-Langevin B\"{o}deker kinetic theory of hot QCD~\cite{Bodeker:1999ey,Litim:1999id} including both noise and collision terms is recovered. In a follow-up work, the formalism developed here will be worked out along similar lines to derive the analogous ``anomalous" B\"{o}deker theory~\cite{Akamatsu:2014yza,Akamatsu:2015kau,follow-up}. As alluded to previously, these can then be matched to results from classical-statistical simulations at early times and to anomalous hydrodynamics at late times.

\section*{Acknowledgments}
RV thanks the Institut f\"{u}r Theoretische Physik, Heidelberg for their kind hospitality and the Excellence Initiative of Heidelberg University for a Guest Professorship during the period when this work was 
initiated; we thank Juergen Berges, Jan Pawlowski, and Michael Schmidt for encouraging this effort. We thank Cristina Manuel and Naoki Yamamoto for valuable comments. We thank Fiorenzo Bastianelli for many helpful discussions and for sharing his deep insights into 
the world-line formulation of quantum field theory. RV would also like to thank the attendees of a seminar on this work at Stony Brook for their helpful comments; in particular, he would like to thank Dima Kharzeev, Ho-Ung Yee, Yi Yin and Ismail Zahed. 

NM acknowledges support by the Studienstiftung des Deutschen Volkes and by the DFG Collaborative Research Centre SFB 1225 (ISOQUANT). This material is partially based upon work supported by the U.S. Department of Energy,
Office of Science, Office of Nuclear Physics, under contract No. DE- SC0012704, and within the framework of the Beam Energy Scan Theory (BEST) Topical Collaboration.


\begin{thebibliography}{90}
\bibitem{footnote1}%
The computations are for over-occupied SU(2) gauge fields in a fixed box. More realistic geometries and $N_c=3$ will not modify this qualitative conclusion.
\bibitem{footnote2}
Couplings are absorbed into the definitions of the fields.
\bibitem{footnote3}
The $8\times 8$ Gamma matrices in $\tilde{\Sigma}^2$ should be understood 
as multiplying the diagonal terms of the $2\times 2$ matrices formed by the gauge fields.
\bibitem{footnote5}
A detailed discussion of the constrained Hamilton dynamics of world-lines can be found in \cite{long-paper}.
\bibitem{footnotepsi0}
The variable $\psi_0$ is not dynamical and we will drop it henceforth.
\end{thebibliography}
\end{document}